\documentclass{IEEEtran}
\usepackage[utf8]{inputenc} 
\usepackage[T1]{fontenc}
\usepackage{url}
\usepackage{ifthen}
\usepackage{cite}
\usepackage{amssymb}
\usepackage{amsfonts}
\usepackage{amsthm}
\usepackage[cmex10]{amsmath}
\usepackage{dsfont}
\usepackage{balance}

\usepackage{graphicx}
\usepackage{hyperref}                                   
\hypersetup{colorlinks=true,allcolors=blue}
\usepackage{xargs}                                      
\usepackage{xcolor}                                     
\usepackage{algorithm}                                  
\usepackage{algpseudocode}                              
\usepackage{authblk}                                    
\usepackage[printonlyused]{acronym}                     
\usepackage{enumitem}                                   
\usepackage{lipsum}                                     

\newtheorem{theorem}{Theorem}

\newtheorem{lemma}{Lemma}

\newtheorem{proposition}{Proposition}

\acrodef{HT}[HT]{Hypothesis Testing}
\acrodef{DHT}[DHT]{Distributed Hypothesis Testing}
\acrodef{SHT}[SHT]{Sequential \ac{HT}}
\acrodef{KLD}[KLD]{Kullback-Leibler Divergence}
\acrodef{DM}[DM]{Decision-Maker}
\acrodef{VM}[VM]{Vending Machine}
\acrodef{SI}[SI]{Side Information}
\acrodef{DMC}[DMC]{Discrete Memoryless Channel}
\acrodef{TAI}[TAI]{Test Against Independence}
\acrodef{PDF}[PDF]{Probability Density Function}
\acrodef{PMF}[PMF]{Probability Mass Function}
\acrodef{CDF}[CDF]{Cumulative Distribution Function}
\acrodef{LRT}[LRT]{Likelihood Ratio Test}
\acrodef{LLR}[LLR]{Log-Likelihood Ratio}
\acrodef{LLRT}[LLRT]{\ac{LLR} Test}
\acrodef{SPRT}[SPRT]{Sequential Probability Ratio Test}
\acrodef{MAP}[MAP]{Maximum A Posteriori}
\acrodef{ML}[ML]{Maximum Likelihood}
\acrodef{BAC}[BAC]{Binary Asymmetric Channel}
\acrodef{UKP}[UKP]{Unbounded Knapsack Problem}
\acrodef{NPT}[NPT]{Neyman-Pearson Test}
\acrodef{AI}[AI]{Artificial Intelligence}
\acrodef{ML}[ML]{Machine Learning}
\acrodef{ABR}[ABR]{Average Bayes Risk}
\acrodef{AEP}[AEP]{Asymptotic Equipartition Property}
\acrodef{MWDT}[MWDT]{Minimal Weight Decision Tree}
\acrodef{TVD}[TVD]{Total Variation Distance}
\acrodef{MGF}[MGF]{Moment-Generating Function}
\acrodef{ISIT}[ISIT]{IEEE International Symposium on Information Theory}
\acrodef{NHSHT}[NHSHT]{Non-Homogeneous \ac{SHT}}
\acrodef{MSPRT}[MSPRT]{Multihypothesis Sequential Probability Ratio Test}

\newif\ifShowProofSkech
\ShowProofSkechtrue

\newif\ifCompileImages
\CompileImagestrue
\CompileImagesfalse 

\newif\ifPaperAward
\PaperAwardtrue
\PaperAwardfalse 

\newif\ifShowComplexityAnalysis
\ShowComplexityAnalysistrue

\newif\ifShowHypothesisEquivFigure
\ShowHypothesisEquivFiguretrue

\newif\ifShowAppendix
\ShowAppendixtrue

\newif\ifReferToAppendix
\ReferToAppendixtrue

\ifCompileImages
    \usepackage{tikz}
    \usepackage{pgfplots} 
    \usepackage{pgfgantt}
    \usepackage{pdflscape}
    \usetikzlibrary{external}
    \pgfplotsset{compat=1.16, plot coordinates/math parser=true}
    \pgfplotsset{every tick label/.append style={font=\footnotesize}}
    \pgfplotsset{every axis/.append style={label style={font=\footnotesize}, width=7cm, height=5.5cm}}
\fi
\newcommandx{\myVec}[2][2=]{{\underline{\smash{#1}}_{#2}}}                              
\newcommandx{\myMat}[2][2=]{{\mathbf{#1}_{#2}}}                                 
\newcommandx{\vectorComponent}[4][3=, 4=]{{[\myVec{#1}_{#3}]_{#2}^{#4}}}        
\newcommandx{\matrixComponent}[4][4=]{{[\myMat{#1}_{#4}]_{#2,#3}}}              
\newcommand{\KLD}[2]{\mathcal{D}_{KL} ( #1 \Vert #2 )}                          
\newcommand{\KLDij}[0]{\KLD{f_i^{a}}{f_j^{a}}}

\newcommand{\expValDist}[2]{\mathbb{E}_{#2}\left[#1\right]}
\newcommand{\prob}[1]{\operatorname{\mathbb{P}}\left( #1 \right)}               
\newcommand{\bigTheta}[1]{\operatorname{\Theta}\left( #1 \right)}               
\newcommand{\hII}{H_i}

\newcommand{\argmax}[2]{\mathop{\operatorname{argmax}}_{#2} #1}

\newcommand{\abs}[1]{ | #1 |}

\newcommandx{\norm}[2][2=2]{\| #1 \|_{#2}}

\newcommand{\probSimplex}[1]{\operatorname{\mathcal{PS}}(#1)}
\newcommand{\probSimplexA}[0]{\probSimplex{\abs{\mathcal{A}}}}

\newif\ifBlind
\Blindtrue
\Blindfalse 

\title{Active Sequential Hypothesis Testing with Non-Homogeneous Costs}
\ifBlind
    \author{%
      \IEEEauthorblockN{Anonymous Authors}
      \IEEEauthorblockA{%
        }
    }
\else
    \author{
        \IEEEauthorblockN{George Vershinin, Asaf Cohen, and Omer Gurewitz}
        
        \IEEEauthorblockA{The School of Electrical and Computer Engineering,
                        Ben-Gurion University of the Negev, Israel
                        \newline
                        georgeve@post.bgu.ac.il, \{coasaf, gurewitz\}@bgu.ac.il}
    }
\fi
\date{}

\begin{document}

\maketitle
\begin{abstract}
    \ifPaperAward
        THIS PAPER IS ELIGIBLE FOR THE STUDENT PAPER AWARD.
    \fi
We study the Non-Homogeneous Sequential Hypothesis Testing (NHSHT), where a single active Decision-Maker (DM) selects actions with heterogeneous positive costs to identify the true hypothesis under an average error constraint \(\delta\), while minimizing expected total cost paid. Under standard arguments, we show that the objective decomposes into the product of the mean number of samples and the mean per-action cost induced by the policy. This leads to a key design principle: one should optimize the ratio of expectations (expected information gain per expected cost) rather than the expectation of per-step information-per-cost (“bit-per-buck”), which can be suboptimal. We adapt the Chernoff scheme to NHSHT, preserving its classical \(\log 1/\delta\) scaling. In simulations, the adapted scheme reduces mean cost by up to 50\% relative to the classic Chernoff policy and by up to 90\% relative to the naive bit-per-buck heuristic.
\end{abstract}

\section{Introduction}
\label{section: introduction}

\ac{HT} is a central tool in statistical inference used to decide whether the data provide sufficient evidence to accept or reject some hypothesis.
This technique has been used for centuries and has been applied across highly diverse domains, such as physics, manufacturing, finance and economics, environmental science, medicine, and its relevance has only grown with the rise of \ac{ML}/\ac{AI} systems that must make reliable, data-driven decisions at scale.
Modern computing systems, including autonomous controllers, anomaly detectors, quality-control pipelines, and networked sensors, continuously detect events or classify states.
Here, a \ac{DM} must select, among competing hypotheses, the one best supported by observed data.
Classical \ac{HT} does so by computing the \ac{LLR} statistic or posterior probability and comparing it to a threshold (e.g., \cite[Theorem.~11.7]{CoverThomas2006}).

In many practical scenarios, data arrives in real-time, and the \ac{DM} can also choose which sample or source to probe next.
Examples include choosing which router to monitor for cyber-attacks or which diagnostic test to run next for a patient.
This naturally leads to \ac{SHT}, whose main appeal is its ability to guarantee the same error rate as a fixed-sample test while enabling early stopping, e.g., \cite{Wald_1945_SHT}.

Typical \ac{SHT} balances two objectives: minimizing average decision error and minimizing detection delay (traditionally measured by the mean sample size).
A large body of work (e.g., \cite{Armitage1950_SHT_MultipleHypotheses, Chernoff1959SequentialHT, Dragalin_etAl_1999_MSPRT_AsympOpt, Dragalin_etAl_2000_MSPRT_MeanSamplesApprox, Cohen_Zhao2015_SHT_AnomalyDetection, Citron_Cohen_Zhao2024_DGF_on_Hidden_Markov_Chains}) analyzes variants of the \ac{MSPRT} and shows that the optimal scaling of sample complexity is logarithmic in $1/\delta$, where $\delta$ is the target average error probability.
Building on variable-length coding ideas in \cite{Burnashev_1946_VLC_SHT}, Naghshvar and Javidi designed \ac{SHT} policies with stronger optimality guarantees, including optimal scaling in the number of hypotheses \cite{Naghshvar_Javidi2013_SHT_DynamicProgramming}.
Other algorithm branches include iterative pruning of inconsistent hypotheses \cite{Gan_Jia_Li2021_Decision_Tree_SHT, vershinin2025multistageactivesequentialhypothesis}.

While the abovementioned works made significant contributions by providing algorithms, studying their scaling laws, and showing performance guarantees, their cost model, where all actions incur the same cost, disregarding their informativeness or nature, is unrealistic.
In practice, samples originating from different sources are associated with different acquisition costs (i.e., non-homogeneous costs) that need not be correlated with or reflect source informativeness, with \emph{wall-clock latency} being a prime example.
In security, querying a router’s queue length is far quicker than running deep packet inspection.
In medicine, ordering a blood test may accelerate the detection of common conditions compared to a stool test, yet delay the identification of rarer ones.
Treating all actions in these examples as equals can result in misallocated efforts (diverting all inbound traffic to deep packet inspection or overwhelming a single laboratory), leading to excessive detection delays.

Although the notion of non-homogeneous costs is not new to some related fields of sequential decision-making, e.g., in the Multi-Armed Bandit problem \cite{Elumar_etAl_2025_MAB_Costs}, Black-Box Optimization \cite{Xie_etAl_2025_BlackBoxOptimization}, or Change Point Detection \cite{Banerjee_Veeravalli_2012_Costly_ChangePointDetection, hou2024robustquickestchangedetection}, the \ac{SHT} literature lacks similar models to our surprise.
Notably, other types of costs have been studied in the context of \ac{SHT}, e.g., action switching costs in \cite{Vaidhiyan_Sundaresan_2015_Switching_Cost, Lambez2022_DGF_wSwitchCost}.
Here, moving between actions incurs an extra penalty that models system inertia.
While relevant when switching itself is costly, these models do not directly account for non-homogeneous acquisition costs that we consider in this work.

A broader perspective on costly information acquisition is given by Caplin and Dean \cite{Caplin_Dean_2015_CostlyInformationAcquisition}, who characterize optimal stochastic information policies.
However, their general framework does not yield concrete scaling laws in $\delta$, which are central in \ac{SHT}.
Unlike \cite{Caplin_Dean_2015_CostlyInformationAcquisition}, we impose no specific structure on the action costs aside from strict positivity.

When each action is associated with its own cost, the \ac{DM} must balance informativeness and the cost it will pay to acquire each sample in order to minimize the total paid cost before delivering its decision.
Thus, diverging from the classic notion of detection delay quantified by the number of samples, we extend the objective function to be the minimization of the expected total paid cost (while adhering to an average error constraint $\delta$).
We refer to this cost-aware formulation as \ac{NHSHT}.

At first glance, the fix seems trivial by normalizing the relevant information metric, e.g., the \ac{KLD} or the mutual information between prior and posterior, by the action cost and running a standard \ac{SHT} scheme.
We show that this “bit-per-buck” heuristic is wrong.
Our key observation is that the correct optimization should be performed on the ratio between the expected number of information bits gained per sample and the expected cost per sample (both under the action selection policy).
Optimizing this ratio accounts for how often each action is taken and how long it actually takes.

Additionally, we show and discuss how to adapt standard \ac{SHT} policies to the \ac{NHSHT} settings \emph{without degrading the classical $\log(1/\delta)$ behavior}, while also improving performance in the finite-regime.
Particularly, we adapt the seminal Chernoff scheme in \cite{Chernoff1959SequentialHT} to \ac{NHSHT}, yielding a simple \ac{NHSHT} baseline algorithm while exemplifying the significance of optimizing the ratio of expectations.

\section{System Model}
\label{section: system model}

\subsection{Notation}
\label{subsection: notation}
All vectors in this manuscript are column vectors and are underlined (e.g., $\myVec{x}$).
The transpose operation is denoted by $(\cdot)^T$.
For the all-zeroes vector, we write $\myVec{0}$, and, similarly, $\myVec{1}$ is the all-ones vector.
Vector components are specified in conjunction with square brackets, e.g., $\vectorComponent{y}{i}$ is $\myVec{y} $’s $i^{\mathrm{th}}$ component.
We write $\myVec{x}\leq \myVec{y}$ as a shorthand notation for $\vectorComponent{x}{i}\leq \vectorComponent{y}{i}$ for all $i$.

The expectation with respect to some random variable $\theta$ is denoted as $\expValDist{\cdot}{\theta}$.
Their \ac{KLD} is denoted as $\KLD{f}{g}$.
Unless explicitly specified (e.g., $\ln$), all logarithms in this manuscript are in base two.
Throughout this paper, we adopt the Bachmann–Landau big-O asymptotic notation as defined in \cite[Chapter~3]{Cormen2009IntroToAlgo3}.

As accepted in the literature, e.g., \cite[Chapter~11]{CoverThomas2006}, discrete distributions over finite alphabets of size $n$ will be regarded as vectors over the probability simplex, $\probSimplex{n} \triangleq \{ \myVec{x}\in\mathbb{R}^n: \myVec{x}^T\myVec{1} = 1, \myVec{x}\geq \myVec{0} \}$.

\subsection{Model}
\label{subsection: model}

The system model consists of a single \ac{DM} capable of obtaining samples from the environment according to the different actions taken from a given set of actions $\mathcal{A} = \{1, 2, \dots, \abs{\mathcal{A}}\}$ with $\abs{\mathcal{A}} < \infty$.
Specifically, at time step $n$, if action $A_n\in\mathcal{A}$ is taken, a \emph{constant} cost $c_{A_n}\in(0, \infty)$ is induced, and the environment outputs a sample $X_n\sim f_\theta^{a}$, where $\theta\in\mathcal{H} = \{0,1,\dots H-1\}$ indicates system state out of $H <\infty$, and $f_\theta^{a}(x) = f(x| a, \theta)$ is the conditional \ac{PDF} of $X_n$ when the underlying system state is $\theta$ under action $a$.
The system state $\theta$ is assumed to have a uniform prior, i.e., $\prob{\theta = i} = \frac{1}{H}$ for any $i$, and we will address the system state as a hypothesis.
Namely, for $\theta = i$, we say that the underlying state follows hypothesis $i$, or $\hII$ for short.

We assume that all obtained samples are conditionally independent.
Extension to non-scalar samples is straightforward and will not be discussed in this work.
All distributions under each hypothesis, i.e., $\{f_h^a\}_{h\in\mathcal{H}, a\in\mathcal{A}}$ are assumed to be known by the \ac{DM}.

\begin{figure}[!htbp]
    \centering
    \includegraphics[]{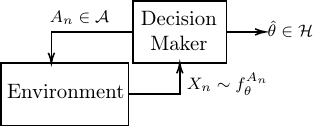}
    \vspace{-8pt}
    \caption{
        System model.
        The \ac{DM} is tasked to identify the correct hypothesis indexed by $\theta\in\mathcal{H}$.
        By taking action $A_n$ at time step $n$, the \ac{DM} obtains a sample $X_n\sim f_\theta^{A_n}$.
        Note that the alphabet of $X_n$ depends on the action $A_n$.
    }
    \label{fig: model}
\end{figure}

The realization of $\theta$ when the \ac{DM} begins operating is unknown, and the \ac{DM}’s goal is to recover $\theta$’s realized value under an average error probability constraint, which will be discussed later.
The final decision made is given by $\hat{\theta}\in\mathcal{H}$, i.e., $\hat{\theta} = i$ implies that $\hII$ is declared as true, or, equivalently, the system state is declared as $i$.
Figure \ref{fig: model} visualizes the model.

We make additional assumptions:
\begin{enumerate}[label=(A\arabic*)]
    \item (Separation) For any action $a\in \mathcal{A}$, for any $i$, $j\in\mathcal{H}$, $\KLDij$ is either 0 or strictly greater than 0.
    \label{assumption: separation}
    \item (Validity) For all $i, j\in\mathcal{H}$ with $i\neq j$, there is some $a\in\mathcal{A}$ with $\KLDij > 0$.
    Furthermore, there is no $a$ with $\KLDij = 0$ for all $i$, $j\in\mathcal{H}$.
    \label{assumption: validity}
    \item (Finite \ac{LLR} Variance) There exists some $\beta>0$ such that $\expValDist{\abs{\log\frac{f_i^a(X)}{f_j^a(X)}}^2 }{f_{i}^a} < \beta$ for any $i, j\in\mathcal{H}$.
    \label{assumption: finite LLR variance}
\end{enumerate}
The Separation and Validity assumptions ensure that there are no meaningless actions and that at least some of the distributions are separated under each action.
The last assumption, first introduced by Chernoff \cite{Chernoff1959SequentialHT}, allows collected \ac{LLR}s to concentrate faster around their mean.

Let $\Phi$ be the source selection process generating the action sequence $\{A_n\}_{n=1}^\infty$.
The source selection rule is \emph{non-adaptive} if the actions do not depend on the gathered data for any time step, and is \emph{adaptive} otherwise.
It may also be either deterministic or stochastic.

Diverging from other works in the literature, we focus on minimizing the total paid cost while adhering to an average error probability constraint less than or equal to any given $\delta$.
Namely, if $\Gamma\triangleq(\Phi, \hat{\theta}, N)$ is an admissible strategy for the \ac{NHSHT} then it is a strategy solving
\begin{equation}
    \label{eq: original problem}
    \begin{split}
        \min_{a_1, a_2, \dots a_N}\quad& \expValDist{\sum_{n=1}^{N} c_{a_n} \middle| \Gamma}{N}
        \\
        \mathrm{s. t.}\quad&    p_e\leq \delta
    \end{split}
\end{equation}
where $N$ is the number of samples upon algorithm termination under policy $\Gamma$, $p_e\triangleq \mathbb{P}(\hat{\theta}\neq \theta | \Gamma)$ is the average error probability of $\Gamma$, and $\hat{\theta}$ is delivered upon termination.
We drop the conditioning on the policy $\Gamma$ to simplify notation.

\subsection{Reviewing the Model and Problem}
\label{subsection: model and problem review}

Assume $\expValDist{N}{N}$ exists and is finite.
The same applies to $\expValDist{N|\theta}{N}$ since $\expValDist{N}{N} = \expValDist{ \expValDist{N | \theta}{N} }{\theta}$ by the Smoothing Theorem \cite[Section~3.4.2]{leongarcia2008}.
Let $N_a$ denote the number of times action $a$ is taken.
Since $\expValDist{N|\theta}{N}<\infty$, then so is $\expValDist{N_a|\theta}{N_a}$ for any $a$.
Hence, the ratio $\expValDist{N_a|\theta = i}{N_a}/\expValDist{N|\theta=i}{N}$ is well defined.
Accordingly, let $\myVec{\lambda}_i \triangleq \{ \expValDist{N_a|\theta = i}{N_a}/\expValDist{N|\theta=i}{N} \}_{a\in\mathcal{A}}$.
Note that $\myVec{\lambda}_i \in \probSimplexA$.
Let $A_i\sim\myVec{\lambda}_i$, and let $A$ be a random variable whose conditional distribution over $\theta=i$ is $\myVec{\lambda}_i$.
Thus,
\begin{align}
    \nonumber
    &\expValDist{\sum_{n=1}^{N} c_{a_n} \middle| \theta = i}{N}
    =
    \expValDist{\sum_{a\in\mathcal{A}} c_{a}N_a \middle| \theta = i }{ N }
    \\
    \nonumber
    &\quad\quad\quad\quad\quad\quad
    =
    \sum_{a\in\mathcal{A}} c_{a}\times\expValDist{N_a | \theta=i}{N_a}
    \\
    \nonumber
    &\quad\quad\quad\quad\quad\quad
    =
    \frac{\expValDist{N|\theta=i}{N}}{\expValDist{N|\theta=i}{N}}\times\sum_{a\in\mathcal{A}} c_{a}\times\expValDist{N_a | \theta=i}{N_a}
    \\
    \nonumber
    &\quad\quad\quad\quad\quad\quad
    =
    \expValDist{N|\theta=i}{N} \times \sum_{a\in\mathcal{A}} c_{a} \times\frac{\expValDist{N_a | \theta=i}{N_a}}{\expValDist{N|\theta=i}{N}}
    \\
    \nonumber
    &\quad\quad\quad\quad\quad\quad
    =
    \expValDist{N|\theta=i}{N} \times \sum_{a\in\mathcal{A}} c_{a} \times\vectorComponent{\lambda}{a}[i]
    \\
    \nonumber
    &\quad\quad\quad\quad\quad\quad
    =
    \expValDist{N|\theta=i}{N} \times \expValDist{c_{A_i}}{A_i}
    \\
    \label{eq: conditional expectation of original}
    &\quad\quad\quad\quad\quad\quad
    =
    \expValDist{N|\theta=i}{N} \times \expValDist{c_{A} | \theta = i}{A}
    .
\end{align}
Consequently, by the Smoothing Theorem,
\begin{align}
    \label{eq: objective function as expectations}
    \expValDist{\sum_{n=1}^{N} c_{a_n}}{N}
    =
    \expValDist{
        \expValDist{N|\theta}{N} \times \expValDist{c_{A}|\theta}{A}
    }{\theta}
    .
\end{align}
We have three important observations;
(i) Optimizing $\expValDist{c_{A}|\theta}{A}$ is not trivial; it is possible that the cheap actions cannot guarantee detection or, alternatively, require an enormous number of samples to separate some hypotheses, with the total cost exceeding the strategy of applying a single, informative yet costly action.
(ii) Since $\expValDist{c_{A}|\theta}{A}$ depends on the underlying policy $\Gamma$, cost-constraint works on variable-length coding, e.g., \cite{Nakiboglu_Gallager_2008_VLC_CostConstrains}, do not directly apply to \ac{NHSHT} as the cost constraint is a function of the algorithm used.
(iii) While the paid cost in practice depends on the particular realization of $\theta$, it is still possible for $\expValDist{c_{A}|\theta}{A}$ to be independent of $\delta$.

Eq. \eqref{eq: objective function as expectations} allows us to comprehend the growth of the objective function in \eqref{eq: original problem} as $\delta\to 0$.
In the classic \ac{SHT}, $\expValDist{c_{A}|\theta}{A}$ is replaced by single cost, say $c$.
Thus, the Lagrangian associated with the optimization problem is $c\expValDist{N}{N} + L\left( p_e - \delta \right)$, whose optimization is equivalent to optimizing
\begin{align}
    \label{eq: ABR}
    \expValDist{N}{N} + L'\times p_e
    ,
\end{align}
where $L' = \frac{L}{c}$.
Eq. \eqref{eq: ABR} is often referred to as the \ac{ABR}.
Several formulations often optimize $\frac{1}{L'}\expValDist{N}{N} + p_e$ instead.
Since $L'$ is independent of the action sequences taken, the rich literature on \ac{SHT} focuses on characterizing $\expValDist{N}{N}$ (as a function of $L'$ or its reciprocal interpreted as $\delta$) for different algorithms or settings.
Typically, $\expValDist{N}{N} = \bigTheta{\log\frac{1}{\delta}}$ with respect to $\delta\to0$, i.e., the number of samples grows logarithmically in $\frac{1}{\delta}$.

On the other hand, the Lagrangian associated with the original formulation in \eqref{eq: original problem} is $\expValDist{N|\theta=i}{N} + \frac{L}{\expValDist{c_{A}|\theta = i}{A}}\times(p_e-\delta)$ for any $i\in\mathcal{H}$.
Namely, the actions taken directly affect the analogous to $L'$ in our problem.
Accordingly, in order to enjoy the optimal scaling laws in $\delta$, \ac{SHT} algorithms must be adapted in a way that $\expValDist{c_{A}|\theta}{A}$ is not a function of $\delta$, and then \eqref{eq: objective function as expectations} is also $\bigTheta{\log\frac{1}{\delta}}$.

To emphasize this point, we revisit the \emph{universal} lower bound in \cite[Proposition~1]{Naghshvar_Javidi2013_SHT_DynamicProgramming} (to be precise, this is Eq. (80) in their supplementary article \cite{Naghshvar_Javidi2013_SHT_DynamicProgramming_Supp} combined with their comment in the discussion in \cite[Section~7.2]{Naghshvar_Javidi2013_SHT_DynamicProgramming}):
\begin{proposition}[Eq. (80) in \cite{Naghshvar_Javidi2013_SHT_DynamicProgramming_Supp}]
    \label{prop: NJ universal LB}
    For any $L'>1$ and arbitrary $\eta\in(0, 1)$, $\expValDist{N}{\theta}$ is lower bounded by
    \begin{align*}
        \frac{1}{H}&\sum_{i=0}^{H-1}\frac{ (1-\eta)\log\frac{1}{\delta} }{ \max_{ \myVec{\lambda}_i\in\probSimplexA }{\min_{j\neq i} \expValDist{ \KLD{f_i^A}{f_j^A} }{A\sim \myVec{\lambda}_i}} +\eta }
        \\
        &\quad\quad\quad\times\left(
            1-2(H-1)H\delta^{\eta}
        \right)
        -(H-1)\beta\eta^{-2}
        .
    \end{align*}
\end{proposition}
Denote $I_i(\myVec{\lambda}) \triangleq \min_{j\neq i} \expValDist{ \KLD{f_i^A}{f_j^A} }{A\sim \myVec{\lambda}}$.
Notably, for sufficiently small $\delta$ and $\eta = \eta(\delta)$ such that $\lim_{\delta\to0}\eta\to 0$ and $\lim_{\delta\to0}\eta^{2}\log\frac{1}{\delta}\to\infty$, the lower bound is dominated by
\begin{align}
    \label{eq: NJ dominating term}
    \frac{1}{H}\sum_{i=0}^{H-1}\frac{ \log\frac{1}{\delta} }{ \max_{ \myVec{\lambda}_i\in\probSimplexA }{ I_i(\myVec{\lambda}_i) }}
    ,
\end{align}
i.e., \eqref{eq: objective function as expectations} has a lower bound dominated by
\begin{align}
    \label{eq: NJ dominating term costs}
    \frac{1}{H}\sum_{i=0}^{H-1}
        \frac{ \log\frac{1}{\delta} }{
            \max_{ \myVec{\lambda}_i\in\probSimplexA } \{I_i(\myVec{\lambda}_i)/\expValDist{c_{A_i} }{A_i\sim\myVec{\lambda}_i}\} }
\end{align}
for sufficiently small $\delta$ and $\eta$ as descried above (see \ifShowAppendix Appendix \ref{subsection: eq: NJ dominating term proof} \else \cite[Appendix~A-A]{vershinin2025activesequentialhypothesistesting} \fi for details).
Thus, to optimize the objective function in \eqref{eq: original problem}, optimizing the ratio between the mean action separation and the mean cost, rather than each alone, is necessary.

\section{The Adapted Chernoff Scheme}
\label{section: Policies}
We take the pioneering work by Chernoff in \cite{Chernoff1959SequentialHT} as an example to showcase how to adapt existing \ac{SHT} algorithms to \ac{NHSHT}.
In the Chernoff scheme, actions at each time step are taken randomly, and the action distribution is guided by the hypothesis whose posterior probability is the highest.
Once the posterior probability of at least one hypothesis exceeds $1-\delta$, the procedure terminates and the \ac{DM} declares the hypothesis with the largest posterior as the true hypothesis.
Specifically, the guiding distribution for $\hII$ is the one optimizing $I_i(\myVec{\lambda})$:
\begin{align}
    \label{eq: Chernoff Policy Dist}
    \myVec{\lambda}_i^*
    =
    \argmax{\min_{j\neq i}\expValDist{ \KLD{f_i^A}{f_j^A} }{A\sim \myVec{\lambda}_i}}{ \myVec{\lambda}_i\in\probSimplexA }
    .
\end{align}
Note that although Chernoff’s model replaced Assumption \ref{assumption: separation} with a stronger assumption: “for any $a\in\mathcal{A}$ and $j\neq i\in\mathcal{H}$, $\KLDij > 0$,” Chernoff’s scheme retains its asymptotic optimality under Assumption \ref{assumption: separation} as argued in \cite[Theorem~2]{Cohen_Zhao2015_SHT_AnomalyDetection}.
We adjust the guiding distribution for $\hII$ in \eqref{eq: Chernoff Policy Dist} to be the distribution optimizing the ratio $I_i(\myVec{\lambda})/\expValDist{c_{A_i}}{A_i\sim\myVec{\lambda}}$: 
\begin{align}
    \label{eq: Chernoff Policy Dist With Costs}
    \myVec{\lambda}_i^*
    =
    \argmax{
        \frac{ \min_{j\neq i}\expValDist{ \KLD{f_i^A}{f_j^A} }{A\sim \myVec{\lambda}_i} }
             { \expValDist{c_{A_i}}{A_i\sim\myVec{\lambda}_i} }
    }{ \myVec{\lambda}_i\in\probSimplexA }
    .
\end{align}
We emphasize that solving \eqref{eq: Chernoff Policy Dist With Costs} is not the same as solving the closely related bit-per-buck variant:
\begin{align}
    \label{eq: Chernoff Policy Bit-per-Buck}
    \argmax{ \min_{j\neq i}\expValDist{ \KLD{f_i^A}{f_j^A}/c_A }{A\sim \myVec{\lambda}_i} }{ \myVec{\lambda}_i\in\probSimplexA }
    .
\end{align}
The solution to \eqref{eq: Chernoff Policy Dist With Costs} is given in the following:
\begin{lemma}
    \label{lemma: adjusted Chernoff dist}
    Let $\myVec{d}_{ij} = \{\KLD{f_i^a}{f_j^a}\}_{a\in\mathcal{A}}$ be a column vector containing the \ac{KLD} between $f_i^a$ and $f_j^a$ for any $a\in\mathcal{A}$.
    Let $\myVec{c} = \{c_a\}_{a\in\mathcal{A}}$ be a column vector containing all costs associated with the actions.
    Then, the solution to \eqref{eq: Chernoff Policy Dist With Costs} is given by $\myVec{\lambda}_i^* = \myVec{y}^*/z^*$ where the pair $\myVec{y}^*$ and $z^*$ are the solutions of:
    \begin{equation}
        \nonumber
        \begin{split}
            \max_{ \myVec{y}\in\mathbb{R}^{\abs{\mathcal{A}}}, z\in\mathbb{R} }\quad&
                \min_{j\neq i} \myVec{d}_{ij}^T\myVec{y}
            \\
            \mathrm{s. t.}\quad&    
                \begin{cases}
                    \myVec{1}^T\myVec{y} -z = 0, &\myVec{c}^T\myVec{y} = 1 \\
                    \myVec{y} \geq \myVec{0}, &z\geq 0
                \end{cases}
        \end{split}
    \end{equation}
\end{lemma}
Note that $\myVec{\lambda}_i^*$ is always well-defined since $z = 0$ is infeasible for any $\myVec{y}$.
Specifically, if $z=0$ then $\myVec{1}^T\myVec{y} = 0$, which in turn implies that $\myVec{y}=\myVec{0}$, violating the constraint $\myVec{c}^T\myVec{y} = 1$.
\begin{IEEEproof}
    The problem in \eqref{eq: Chernoff Policy Dist With Costs} can be transformed into the above optimization problem in a similar way that Linear-Fractional Programming problems can be transformed to Linear Programming problems, cf. \cite[Chapter~4.3.2]{boyd_vandenberghe_2004}.
    See \ifShowAppendix Appendix \ref{subsection: lemma: adjusted Chernoff dist proof} \else \cite[Appendix~A-B]{vershinin2025activesequentialhypothesistesting} \fi for details.
\end{IEEEproof}

Since the only changed parameter in the adapted Chernoff scheme is the distribution from which actions are taken, its relevant upper and lower bounds on $\expValDist{N}{N}$ still apply.
Particularly, the lower bound in Proposition \ref{prop: NJ universal LB} holds and the upper bound is established from \cite[Lemma~2]{Chernoff1959SequentialHT} which states that for any $\epsilon > 0$ there exists some $\delta_0$ such that for any $\delta \leq \delta_0$ we have $\expValDist{N}{N} \leq (1+\epsilon)\times\frac{1}{H}\sum_{i=0}^{H-1}\frac{\log\frac{1}{\delta}}{ I_i(\myVec{\lambda}_i) }$.
Consequently, for any $\epsilon > 0$ there exists some $\delta_0$ such that for any $\delta \leq \delta_0$:
\begin{align}
    \label{eq: Chernoff eventual UB}
    \eqref{eq: objective function as expectations} \leq \expValDist{c_{A}}{A}\times(1+\epsilon)\times\frac{1}{H}\sum_{i=0}^{H-1}\frac{\log\frac{1}{\delta}}{ I_i(\myVec{\lambda}_i) }
    ,
\end{align}
where $\expValDist{c_{A}}{A} = \expValDist{\expValDist{c_{A}|\theta}{A}}{\theta}$ from the Smoothing Theorem.
Accordingly, the following result directly obtained by combining Proposition \ref{prop: NJ universal LB} and Eq. \eqref{eq: Chernoff eventual UB}:
\begin{theorem}[Cost-Aware Chernoff Scheme]
    \label{theorem: Chernoff order optimality}
    Let $\myVec{\rho}_n\in\probSimplex{H}$ be the posterior probability over the hypotheses.
    That is, $\vectorComponent{\rho}{i}[n] = \vectorComponent{\rho}{i}[n-1]\times f_i^a(x)/(\sum_{h\in\mathcal{H}} \vectorComponent{\rho}{h}[n-1]\times f_h^a(x))$.
    Let $\{\myVec{\lambda}_h^*\}_{h\in\mathcal{H}}$ be solutions for \eqref{eq: Chernoff Policy Dist With Costs} obtained via Lemma \ref{lemma: adjusted Chernoff dist}.
    The policy taking $A_n\sim \myVec{\lambda}_{ h_n^* }^*$ where $h_n^* = \argmax{ \vectorComponent{\rho}{h}[n] }{ h\in\mathcal{H} }$ until some posterior exceeds $1-\delta$ for the first time ensures that \eqref{eq: objective function as expectations} is $\bigTheta{\log\frac{1}{\delta}}$ (with respect to $\delta\to 0$) and $p_e \leq \delta$.
\end{theorem}

\section{Numerical Results}
\label{subsection: Numerical Results}
In this section, we present simulation results illustrating our findings.
The number of hypotheses was set to $H=32$.
For simplicity, all $\abs{\mathcal{A}} = 16$ actions produce unit-variance normally distributed samples, where half of the hypotheses are assigned a mean of $2$, and the others a mean of $8$ at random.
Then, each mean is contaminated by uniform $[-0.1, 0.1]$ noise.
Note that Assumption \ref{assumption: finite LLR variance} holds with $\beta = 361$.
Finally, $H_0$ and $H_{31}$ had their means set to be the same for all actions (enforcing Assumption \ref{assumption: separation}) but the last, wherein $\mu_0 = 10-\mu_{31}$, so Assumption \ref{assumption: validity} holds.
We emphasize that the means are drawn only once and remain fixed throughout the simulation.
The cost vector is $\myVec{c}=(1, 10, 20, 30, \dots, 150)^T$, i.e., action one costs 1, action two costs 20, etc.

\begin{figure}[!htbp]
    \centering
    \includegraphics[scale=0.8]{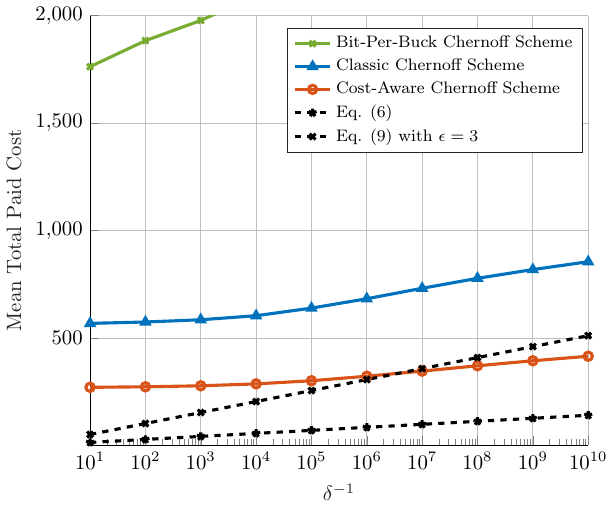}
    \vspace{-8pt}
    \caption{
        The expected total paid costs of the classic Chernoff and Cost-Aware Chernoff schemes.
        All samples are unit-variance, normally distributed with randomly drawn means.
    }
    \label{figure: expected total paid cost vs. delta norm}
\end{figure}

The simulation results are presented in Figure \ref{figure: expected total paid cost vs. delta norm}.
Here, the green, blue, and orange curves are the expected total paid costs (the objective function in \eqref{eq: original problem}) averaged over $50000$ different iterations of different Chernoff schemes.
The orange curve is the Cost-Aware Chernoff scheme (Theorem \ref{theorem: Chernoff order optimality}) whose guiding distributions are computed via Lemma \ref{lemma: adjusted Chernoff dist}.
In blue, we present the classic Chernoff scheme’s expected total paid cost, whose guiding distribution for $\hII$ is the one obtained by solving \eqref{eq: Chernoff Policy Dist}.
The green curve is the “bit-per-buck” Chernoff scheme guided by the solution to \eqref{eq: Chernoff Policy Bit-per-Buck}, whose smallest expected total paid cost is $\sim$1762.318 and monotonously grows up to $\sim$4700.
The latter should be compared to the maximal paid cost of $\sim$855 by the classic Chernoff Scheme or the $\sim$415 by the Cost-Aware Chernoff scheme.

The dashed black curves are the expression in \eqref{eq: NJ dominating term costs} and \eqref{eq: Chernoff eventual UB}, for which we took $\epsilon = 3$ so the upper bound will hold for some large $\delta_0\in[10^{-10}, 10^{-1}]$.
The expected total paid cost by the Cost-Aware Chernoff scheme eventually (i.e, for any $\delta\leq\delta_1$ for some $\delta_1$) has the same slope as the lower bound in Eq. \eqref{eq: NJ dominating term costs}.
This observation validates the importance of optimizing the ratio of expected informativeness and expected cost. 

\section{Conclusion}
\label{section: conclusion}

We introduced the \ac{NHSHT}, a cost-aware variant of \ac{SHT} in which actions carry non-homogeneous positive costs.
We showed that minimizing expected total cost is followed by optimizing the ratio between the expected number of information bits gained per sample and the expected cost per sample (both conditioned on the action selection policy itself).
Building on this principle, we adapted the Chernoff scheme to \ac{NHSHT} while retaining its scaling laws.
Empirically, the adapted policy outperforms both the classic Chernoff scheme and the “bit-per-buck” heuristic, reducing mean total paid cost by 50\% and by an order of magnitude, respectively.


\clearpage
\balance
\bibliographystyle{IEEEtran}
\bibliography{references}

\ifShowAppendix
    \appendices
\section{Miscellaneous Proofs}
\label{section: misc proofs}

\subsection{Bounding The Objective Function Using \texorpdfstring{\eqref{eq: NJ dominating term}}{NJ LB dominating term}}
\label{subsection: eq: NJ dominating term proof}

Here, we leverage the conditional expectation in \eqref{eq: conditional expectation of original} and combine it with \eqref{eq: NJ dominating term}.
Specifically, each addend inside the sum in \eqref{eq: NJ dominating term} bounds $\expValDist{N | \theta=i}{N}$, so
\begin{align*}
        \eqref{eq: conditional expectation of original}
        &\geq
        \frac{ \expValDist{c_{A} | \theta = i}{A} \times \log\frac{1}{\delta} }
             { \max_{ \myVec{\lambda}_i\in\probSimplexA }{\min_{j\neq i} \expValDist{ \KLD{f_i^A}{f_j^A} }{A\sim \myVec{\lambda}_i}} }
        \\
        &
        =
        \frac{ \expValDist{c_{A} | \theta = i}{A} \times \log\frac{1}{\delta} }
             { \max_{ \myVec{\lambda}_i\in\probSimplexA }{ I_i(\myVec{\lambda}_i) } }
        \\
        &
        =
        \frac{ \log\frac{1}{\delta} }
             { \frac{ \max_{ \myVec{\lambda}_i\in\probSimplexA }{ I_i(\myVec{\lambda}_i) } }
                    { \expValDist{c_{A} | \theta = i}{A} } }
        \\
        &
        =
        \frac{ \log\frac{1}{\delta} }
             { \frac{ \max_{ \myVec{\lambda}_i\in\probSimplexA }{ I_i(\myVec{\lambda}_i) } }
                    { \expValDist{c_{A_i}}{A_i\sim\myVec{\lambda}_i} } }
        \\
        &
        \geq
        \frac{ \log\frac{1}{\delta} }
             { \max_{ \myVec{\lambda}_i\in\probSimplexA } \frac{ I_i(\myVec{\lambda}_i) }
                                                               { \expValDist{c_{A_i}}{A_i\sim\myVec{\lambda}_i} } }                      
\end{align*}

\subsection{Proof of Lemma \ref{lemma: adjusted Chernoff dist}}
\label{subsection: lemma: adjusted Chernoff dist proof}

With the new notation, \eqref{eq: Chernoff Policy Dist With Costs} can be rewritten as
\begin{equation}
    \label{eq: Chernoff Policy Dist With Costs 2}
    \begin{split}
        \max_{ \myVec{\lambda}_i\in\mathbb{R}^{\abs{\mathcal{A}}} }\quad&
            \frac{ \min_{j\neq i} \myVec{d}_{ij}^T\myVec{\lambda}_i }
            { \myVec{c}^T\myVec{\lambda}_i }
        \\
        \mathrm{s. t.}\quad&    
            \begin{cases}
                \myVec{1}^T\myVec{\lambda}_i = 1 \\
                \myVec{\lambda}_i\geq \myVec{0}
            \end{cases}
    \end{split}
    .
\end{equation}
We now show equivalence.
For any $\myVec{\nu}$ feasible in \eqref{eq: Chernoff Policy Dist With Costs 2}, the pair $\myVec{y} = \myVec{\nu}/(\myVec{c}^T\myVec{\nu})$ and $z = 1/(\myVec{c}^T\myVec{\nu})$ is feasible for the optimization problem in Lemma \ref{lemma: adjusted Chernoff dist}.
The value of the objective function in the problem in Lemma \ref{lemma: adjusted Chernoff dist} becomes $\min_{j\neq i} \myVec{d}_{ij}^T\myVec{\nu} /(\myVec{c}^T\myVec{\nu})$ in this case, which is the same value the objective function in \eqref{eq: Chernoff Policy Dist With Costs 2} obtains when setting $\myVec{\lambda}_i = \myVec{\nu}$.

Assume $\myVec{y}$ and $z$ are feasible for the problem in Lemma \ref{lemma: adjusted Chernoff dist}.
Note that $z = \myVec{1}^T \myVec{y} \neq 0$ (otherwise, $\myVec{y}=\myVec{0}$ and the constraint $\myVec{c}^T\myVec{y} = 1$ does not hold).
We observe that $\myVec{\nu} = \myVec{y}/z\geq\myVec{0}$ is feasible for \eqref{eq: Chernoff Policy Dist With Costs 2} as $\myVec{1}^T\myVec{\nu} = \myVec{1}^T \myVec{y}\times\frac{1}{z} = \myVec{1}^T \myVec{y} \times (\myVec{1}^T \myVec{y})^{-1} = 1$.
Particularly, setting $\myVec{\lambda}_i = \myVec{y}/z$ inside the objective function in \eqref{eq: Chernoff Policy Dist With Costs 2} results in the value of $\min_{j\neq i} \myVec{d}_{ij}^T\myVec{y}$, which is the same value as of the objective function in Lemma \ref{lemma: adjusted Chernoff dist}.


\fi

\end{document}